\documentstyle[aps,prb,eqsecnum,psfig,epsf,floats]{revtex}
\begin{document}
\twocolumn[\hsize\textwidth\columnwidth\hsize\csname %
@twocolumnfalse\endcsname
\title{Phenomenological model for magnetotransport in a
multi-orbital system}
\author{Canio Noce and Mario Cuoco \\
I.N.F.M. -Unit\`a di Salerno, Dipartimento di Scienze Fisiche
''E.R. Caianiello'',\\ Universit\`a di Salerno \\ I-84081
Baronissi (Salerno), ITALY} \maketitle

\begin{abstract}
\noindent By means of the Boltzmann equation, we have calculated
some magnetotransport quantities for the layered multi-orbital
compound Sr$_2$RuO$_4$. The Hall coefficient, the
magnetoresistance and the in-plane resistivity have been
determined taking into account the Fermi surface curvature and
different time collisions for the electrons in the $t_{2g}$
bands. A consistent explanation of the experimental results has
been obtained assuming different relaxation rates for the
in-plane transport with and without an applied magnetic field,
respectively.
\end{abstract}
] \narrowtext
\bigskip

The layered perovskite oxide Sr$_2$RuO$_4$ has attracted
considerable experimental and theoretical attention since the
recent discovery of superconductivity in this
compound.{\cite{maeno}} Quantitative similarities between the
Fermi liquid in $^3$He and that in Sr$_2$RuO$_4$ hint at the
possibility of p-wave superconducting pairing.{\cite{rice}} A
growing body of experimental evidence, including results obtained
from muon spin relaxation,{\cite{luke}} NMR 1/T$_1$ and Knight
shift,{\cite{ishida}} neutron scattering,{\cite{rise}} impurity
effect,{\cite{andy2}} and specific heat measurements,{\cite{CV}}
has shown that the pairing symmetry is unconventional, most likely
p-wave.

This compound shows also interesting normal state transport
properties characterized by multi-bands effects, and provides an
ideal opportunity to investigate the crossover from standard to
non standard conduction processes. The in plane zero-field
resistivity exhibits a Fermi-liquid like behaviour up to
T$_{FL}\equiv $25 K; above this temperature $\rho_{ab}$ rises
monotonically with a curvature which is much weaker than
T$^2$.\cite{mac2} Moreover, around 1300 K the in-plane mean free
path falls smoothly to less than $1 \AA$ with no sign of
resistivity saturation at the Mott-Ioffe-Regel limit, clearly
indicating an unconventional high temperature conducting
mechanism.\cite{TMNM}

$\rho _c$ reaches a maximum at 130 K then slowly decreases with
increasing temperature.{\cite{hus}} Below $\sim$ T$_{FL}$ the
out-of -plane resistivity has a quadratic temperature dependence
indicating that a crossover to a Fermi-liquid state takes place
below 25K prior to the superconducting transition. Thus, below
T$_{FL}$ Sr$_2$RuO$_4$ has a resistivity with a quadratic
temperature dependence in all directions and it is extremely
anisotropic, with $\rho _c/$ $\rho _{ab}>$500, a ratio that is
observed in most of the cuprate materials.

The Hall coefficient exhibits a complicated temperature
behavior:{\cite{mac2}} below 1 K it becomes almost temperature
independent assuming a value of -1.15$\times$10$^{-10}$m$^3$/C; it
has a strong temperature dependence below 25 K then, changes sign
at temperatures approximately above $30$ K and shows a return to
negative values above approximately $130$ K.

The $c-$axis magnetoresistance{\cite{hus}} is large and positive
and varies linearly with the applied magnetic field; with the
increase of the temperature it falls sharply becoming negative
above $75$ K. The in-plane magnetoresistance is positive and large
at low temperatures, then decreases as T is raised up to $80$ K.

The aim of the present work is to show, by means of the Boltzmann
equation, that an overall agreement for the in-plane
magnetotransport quantities can be reproduced in a temperature
range up to 150 K, where the relaxation time approximation is
applicable, i.e. the mean free path is greater than the lattice
spacings. In order to determine the magnetotransport quantities,
one has to calculate the conductivity tensor up to second order
and this in turn implies the knowledge of the energy band
spectrum as well as the explicit expression for the relaxation
rate.

The information of the electronic structure is given in Ref.
{\cite{NOI}} for the three bands crossing the Fermi level and the
loss of coherence of the conduction electrons is simulated by
assuming an ad hoc temperature behavior for the scattering rates
for electrons belonging to different bands. More precisely, the
key assumption is that the temperature behaviour of the scattering
rate of the carriers is modified by the presence of an applied
magnetic field. We argue that the relaxation rate produced by spin
fluctuations is suppressed in presence of an external magnetic
field implying that the main contribution comes only from the
usual electron-electron, impurity, or phonon- mediated scattering,
which sum up to be $\sim$ T$^2$ at low temperatures. This effect
applies in determining the Hall- and magneto- resistance.

On the other hand, in the absence of magnetic field, according to
the Matthiessen rule, the induced spin-fluctuation scattering rate
adds to the previous relaxation mechanisms in determining the
total resistivity of the system.

The presence of an additional scattering rate is supported by the
experimental evidence that the Sr$_2$RuO$_4$ is close to magnetic
instabilities\cite{imai,sidis} which are the source of enhanced
spin fluctuations at different $q$ points of the Brillouin zone,
thus giving rise to large scattering amplitudes between charge
carriers and the spin fluctuations themselves.

$^{17}$O NMR measurements\cite{imai} probe spin correlations in Ru
d$_{xy}$ and d$_{xz,yz}$ orbitals separately and show that only
$\chi _{xy}$ increases monotonically with decreasing temperature
down to about 40K, following a Curie-like behaviour, then turns
over and tends to level below T$_{FL}$, implying that the spin
correlations in the d$_{xy}$ band are predominantly ferromagnetic
in origin.

Furthermore, by comparing $^{101}1/T_1 T$ at the Ru site and
$^{17}1/T_1 T$ at the planar O site, due to the different
dependence of their hyperfine form factor in k-space, one can
probe whether the in-plane spin correlations are ferromagnetic or
antiferromagnetic. It turns out\cite{imai} that both $^{101}1/T_1
T$ and $^{17}1/T_1 T$ increase monotonically down to $T_{FL}$, and
almost saturate in a Korringa-like behaviour.

Moreover, inelastic neutron scattering measurements\cite{sidis} in
the normal state reveal the existence of incommensurate magnetic
spin fluctuations located at ${\bf q_0}=(\pm 0.6 \pi /a,\pm 0.6
\pi /a,0)$ due to the pronounced nesting properties of the almost
one-dimensional $d_{xz,yz}$ bands. In fact, the 1D sheets can be
schematically described by parallel planes separated  by
$\check{q}=\pm 2 \pi /3a$, running both in the $x$ and in the $y$
directions which give rise to dynamical nesting effects at the
wave vectors ${\bf k}=(\check{q},k_y)$, ${\bf k}=(k_x,\check{q})$,
and in particular at ${\bf
\check{q}}=(\check{q},\check{q})$.\cite{Maz99}

Though not as evident as in the $\chi_{xy}$ and the NMR
measurements, due to the presence of few experimental points, the
temperature dependence of the imaginary part of the susceptibility
$\chi^{\prime\prime}$, at low energy and at wave vector
${\bf{q_0}}$, exhibits a sharp decrease upon temperature increase
above a temperature of $\sim$ T$_{FL}$.\cite{sidis}

Combining the results of NMR, nuclear spin-lattice relaxation
rate, and inelastic neutron scattering measurements, it is
possible to draw the following physical picture for the
Sr$_2$RuO$_4$. There is a strong enhancement of spin fluctuations
above T$_{FL}$, mainly due to ferromagnetic correlations between
the electrons in the $d_{xy}$ band, as revealed from Knight shifts
experiments and nuclear spin relaxation rate, and due to
incommensurate contributions from the nesting properties of the
almost 1D $d_{xz,yz}$ bands. Hence, the main features of the
magnetic response turn out to be decoupled for the electrons in
the $d_{xy}$ and in the $d_{xz,yz}$ bands respectively.

We consider that the same decoupling manifests in the transport
properties.

One important experimental observation is that there is a close
relation between the change in the magnetic response and the
resistivity measurements. Below $\sim$T$_{FL}$ the in-plane
resistivity have a quadratic temperature dependence, while above
$\sim$ T$_{FL}$ a superlinear term adds to the T$^2$ contribution
\cite{mac2}.

We assume that above $\sim$ T$_{FL}$ the linear contribution in
the in-plane resistivity is mainly determined by small momentum
scattering in the d$_{xy}$ band derived by the ferromagnetic spin
fluctuations, and we consider also the slow temperature variation
of the relaxation rates ($\sim~T~\ln T$) which comes from the
scattering at large momentum transfer due to the incommensurate
spin fluctuations for the d$_{xz,yz}$ electrons. It is worthwhile
pointing out that according to our calculations, the latter does
not give any substantial qualitative and quantitative change in
the transport properties, mainly due to the limited phase space
allowed for the scattering processes.

Specifying these considerations to the multi-band system in
question, we assume only for the $\gamma$-band, a scattering rate
proportional to T$^2$ in calculating the Hall coefficient and the
magnetoresistance, with an additional term reproducing the effects
of scattering by ferromagnetic spin fluctuations, to determine the
zero-field resistivity.

For a 2D system, the temperature dependence of the scattering rate
due to this mechanism is given by\cite{beal}
\begin{eqnarray*}
\tau^{-1}_{sf} \sim \int q^2 dq \int Im \chi(q,\omega)
\frac{\partial}{\partial T} n(\omega),
\end{eqnarray*}
\noindent where $n(\omega)$ is the Bose distribution and
$\chi(q,\omega)$ is the dynamical susceptibility. Within the
self-consistent spin fluctuation theory, in a paramagnet close to
a ferromagnetic instability, one gets $\tau^{-1}_{sf} \sim
T$.\cite{Moriya}

For the other two bands, a quadratic dependence of the scattering
rates on the temperature for all the calculated quantities is
assumed.

To find the transport coefficients, we must calculate the current
defined as:
\begin{eqnarray}
{\bf J}=\int e {\bf v}g({\bf v}) d{\bf k}.
\end{eqnarray}
where $g({\bf v})$ is the local distribution of electrons.

Confining ourselves to an expansion of order $B^2$, one can easily
obtain the following general formula \cite{ziman}
\begin{eqnarray}
J_\alpha=\sigma^{(0)}_{\alpha \beta} E_{\beta}
+\sigma^{(1)}_{\alpha \beta \gamma} E_{\beta} B_{\gamma}+
\sigma^{(2)}_{\alpha \beta \gamma \delta} E_{\beta} B_{\gamma}
B_{\delta},
\end{eqnarray}
where the summation convention over the indices of the cartesian
components is assumed and $E_{i}$ and $B_{i}$ are the components
of the external electric and magnetic field, respectively.

To calculate the normal resistivity, the Hall coefficient and the
magnetoresistance for a multi-band case, we write the total
current as the sum of the contributions coming from the three
bands and then we invert the matrix connecting ${\bf J}$ and
${\bf E}$. Neglecting powers above $B^2$, we have:
\begin{eqnarray}
\rho_0=\frac{1}{\sigma^{Tot}_0}\quad,
\end{eqnarray}
\begin{eqnarray}
\rho_H=-\rho^2_0 \sigma^{Tot}_H \quad,
\end{eqnarray}
\begin{eqnarray}
\rho_{MR}=-\rho^2_0 \left[
\sigma^{Tot}_{MR}+\rho_0(\sigma^{Tot}_H)^2 \right]\quad ;
\end{eqnarray}
where $\sigma^{Tot}_0$, $\sigma^{Tot}_H$ and $\sigma^{Tot}_{MR}$
denote the total conductivity, the total Hall conductivity and
the second order total conductivity, respectively.

The explicit computation of the magnetotransport quantities
requires the knowledge of the band spectra as well as the
relaxation times for describing the collision of the electrons in
the bands produced by the d$_{xy}$, d$_{xz}$ and d$_{yz}$ Ru
orbitals.

Referring to the energy spectra, we use the electronic energy band
structure of Sr$_2$RuO$_4$ recently calculated by using a simple
method combining the extended H\"uckel theory and the
tight-binding approximation.\cite{NOI}

Concerning the relaxation times, in the case of the Hall- and
magneto- resistance the following expressions have been
considered:
\begin{eqnarray*}
(\tau_{i})^{-1}&=&\eta_{i}+\alpha_{i} T^2
\end{eqnarray*}
where $i$=$(xz,yz,xy)$ indicates the band and
$\eta=(2.75,2.75,3.25)$, and $\alpha=(0.035,0.04,0.06)$. The
values of $\eta_{i}$ have been chosen in a way to get the
experimental observed resistivity at T=4 K of $\sim 0.7 \mu\Omega
cm^{-1}$. The constraint on the values of $\alpha_{i}$ is given by
the complicated temperature dependence of $R_H$ together with the
behaviour of the transverse in-plane magnetoresistance. We notice
that the behaviour of the $R_H$ can be reproduced only if
$\alpha_{xz}\sim \alpha_{yz}$ but smaller than $\alpha_{xy}$. The
calculation is very sensitive to the changes in the time
collisions of the two hybridized $z$ bands. Indeed, for small
relative variation of $\alpha_{xz}$ with respect to $\alpha_{yz}$
the Hall-coefficient does not show sign changing, being always
negative.

It is worth pointing out that the explicit expression for the
scattering rate is related to several physical mechanisms that
give rise to different temperature dependencies of $\tau_i$. In
particular, while $\eta_i$ could be considered as responsible for
the residual resistivity, $\alpha_i$ could be related to umklapp
electron-electron scattering, or inelastic scattering of the
electrons by impurities, as well as by phonon mediated interaction
between electrons. In the case of electron-electron scattering, an
estimation of $\alpha_i$ can be obtained by means of the relation
$(\tau^{e-e}_i)^{-1}\approx (k_B T^2/m_i k^i_F)$. Using the
experimental values for the effective masses as deduced from de
Haas-van Alphen experiment, we find that $\alpha_{xy}$ is greater
than $\alpha_{xz}$ and $\alpha_{yz}$ and this agrees with our
assumption.

Concerning the other mechanisms, though the microscopic expression
in the case of the phononic and impurity scattering requires a
more accurate analysis, we expect that the considerations above
are still valid.

The fit to the experimental data is reported in Fig.\ref{figTR1},
for the Hall coefficient and in Fig.\ref{figTR2} for the
magnetoresistance. The experimental data are taken from
\cite{mac2} for $R_H$ and from \cite{hus} for the
magnetoresistance. In both cases, we find a good agreement between
the experimental results and the theoretical prediction indicating
that the main contribution to scattering rate follows a $T^2$
power law.

We notice that paramagnetic and ferromagnetic materials can have a
large contribution to the Hall effect mainly due to skew
scattering. Indeed, in this case moving charge carriers experience
a force due to the magnetic field produced by localized magnetic
moment and are scattered asymmetrically. Nevertheless, there is no
sign of the saturation of the Hall resistivity one expects when
there is a large magnetic contribution to the Hall
effect.\cite{mac2} Therefore, we argue that the experimental data
for Sr$_{2}$RuO$_{4}$ are dominated by the standard orbital Hall
resistivity.

We point out that there are other results in literature dealing
with the theoretical study of the Hall coefficient in the
Sr$_2$RuO$_4$. In Ref. \cite{Shr95} is presented a theoretical fit
of the Hall coefficient based on the assumption of two-carrier
system and $R_H$ is calculated within the Drude classical model.
While, the authors neglect the contribution of one of the electron
pockets and the effective electronic structure of the bands, they
reproduce the sign change from negative at low temperature to
positive at high temperature but fails to give a quantitative
agreement with the experimental data.

In Ref. \cite{mac2}, using methods developed by Ong, an expression
for the Hall coefficient in multi-band system is derived. We
notice that the main assumption of this derivation is that the
mean free path is the same for all the Fermi sheets. This
hypothesis of isotropic mean-free path is valid at small
temperature where the authors obtain a value for $R_H$ that
compares well with the measured value of the same quantity.
Nevertheless, we want to stress that the value of $R_H$ is
extremely sensitive to details of the k-dependent scattering
and/or on the energy spectra, and strongly depends on the
temperature.

Finally, in Ref. \cite{NOI} is shown a fit to $R_H$ similar to the
one here presented. However, in Ref. \cite{NOI}, the sign change
is reproduced with an accuracy weaker than the one presented in
the present paper and no mention to other relevant physical
quantities is made.
\begin{figure}[h]
\centerline{\psfig{figure=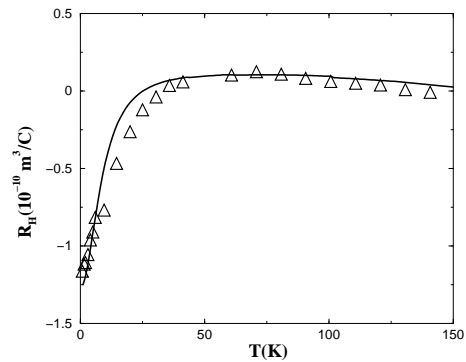,width=6.cm}} \caption{{The
temperature dependence of the weak-field Hall coefficient. The
triangles show data taken from {\protect\cite{mac2}}, and the
solid line is the theoretical result.}} \label{figTR1}
\end{figure}
\begin{figure}[h]
\centerline{\psfig{figure=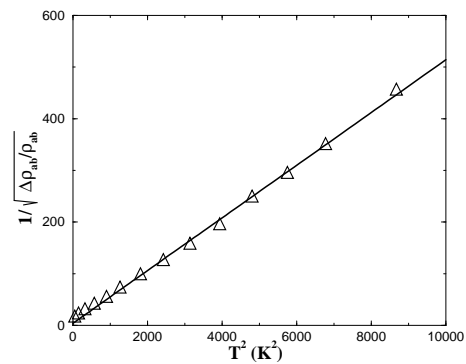,width=6.cm}} \caption {The
temperature dependence of $1/\sqrt{\Delta \rho_{ab}/\rho_{ab}}$
(triangles){\protect\cite{hus}} and the theoretical result (solid
line).} \label{figTR2}
\end{figure}
\begin{figure}
\centerline{\psfig{figure=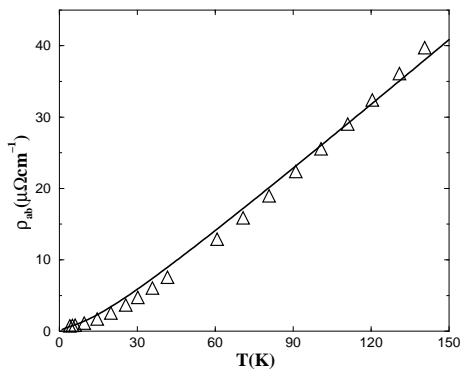,width=6.cm}} \caption{In plane
resistivity vs temperature: the triangles are the experimental
results from  \protect\cite{hus} while the solid line is the
theoretical curve.} \label{figTR3}
\end{figure}

With respect to the previous approaches, therefore, we believe the
results presented here yield a good qualitative and quantitative
agreement  with the transport experiments.

Let us now discuss the zero-field in plane resistivity.

As mentioned in the introduction, $\rho_{ab}$ exhibits a
Fermi-liquid behaviour up to T$_{FL}$; above this temperature
$\rho_{ab}$ rises monotonically with a curvature which is weaker
than T$^2$. We notice that this effect may be probably induced by
a spin scattering mechanism which does not contribute to the
determination of the Hall coefficient and the magnetoresistance.
Therefore, we add to the scattering rate for the $\gamma$ band
electrons above, a term proportional to the temperature, while we
keep unchanged the other two scattering rates. Its temperature
dependence is given by:
\begin{eqnarray*}
(\tau^{sf}_{xy})^{-1}&=&\beta T
\end{eqnarray*}
\noindent where $\beta=0.6$. The resulting fit is reported in
Fig.\ref{figTR3} where the experimental data are taken from
\cite{hus}. The quite good agreement between the experimental
results and the theoretical prediction gives confidence that spin
fluctuations affects only the zero field relaxation time for the
xy band.

As a final remark, we notice that throughout this paper we have
assumed for the relaxation rates an isotropic k-independent form.
We have also evaluated the above mentioned quantities assuming
for $\tau_i$ the suitable form in the case of a tetragonal
environment. The results are only slightly modified, so that we
have confined ourselves to temperature dependent but
k-independent $\tau_i$.

In summary, we have studied the in-plane normal state
magnetotransport quantities of the layered compound
Sr$_2$RuO$_4$. Using the calculated electron energy band
structure, we have computed the temperature dependence of the
Hall coefficient, the magnetoresistance and the in-plane
resistivity by solving the Boltzmann equation for a multi-orbital
system. The reasonably good fit of these physical quantities
suggests that the assumption of two contributions in the
relaxation rate for the xy-electrons used to quantify the
galvanomagnetic transport is essentially correct.

\end{document}